\documentclass[longauth]{aa}  

\usepackage{graphicx}
\usepackage{txfonts}
\usepackage{xcolor}
\usepackage{hyperref}
\usepackage{multirow}
\newcommand{\Ha}{\hbox{H$\alpha$}\xspace}
\newcommand{\Hb}{\hbox{H$\beta$}\xspace}
\newcommand{\Pa}{\hbox{Pa$\alpha$}\xspace}
\newcommand{\OII}{\hbox{[O\,\textsc{ii}]$\,\lambda\lambda3726,3729$}\xspace}
\newcommand{\OIIIa}{\hbox{[O\,\textsc{iii}]$\,\lambda4959$}\xspace}
\newcommand{\OIIIb}{\hbox{[O\,\textsc{iii}]$\,\lambda5007$}\xspace}
\newcommand{\NIIb}{\hbox{[N\,\textsc{ii}]$\,\lambda6584$}\xspace}
\newcommand{\ergscm}{\hbox{$\mathrm{erg\,s^{-1}\,cm^{-2}}$}\xspace}
\newcommand{\Msunyr}{\hbox{$M_{\odot}\,\text{yr}^{-1}$}\xspace}

\newcommand{\zgrb}{\hbox{$z_{\rm GRB} = 0.15095\pm 0.00005$}}
\newcommand{\zem}{\hbox{$z = 0.1510$}\xspace}
\newcommand{\Avhost}{\hbox{$A_V^{\rm host} = 0.4^{+0.2}_{-0.2}$}\xspace}
\newcommand{\SFRhost}{\hbox{${\rm SFR} = 0.16^{+0.02}_{-0.02}$}\xspace}
\newcommand{\logOHhost}{\hbox{${\rm 12 + \log(O/H)} < 8.5$}\xspace}

\begin{document}

   \title{The brightest GRB ever detected: GRB\,221009A as a highly luminous event at $z=0.151$}

\author{D.~B. Malesani\inst{1,2,3}
    \and A.~J. Levan\inst{1,4}    
    \and L. Izzo\inst{5} 
    \and A. de Ugarte Postigo\inst{6}
    \and G. Ghirlanda\inst{7,8}
    \and K.~E. Heintz\inst{2,3}
    \and D.~A. Kann\inst{9}\thanks{Deceased}
    \and G.~P. Lamb\inst{10}
    \and J. Palmerio\inst{11}
    \and O.~S. Salafia\inst{7}
    \and R. Salvaterra\inst{12}
    \and N.~R. Tanvir\inst{13}
    \and J. F. Ag\"u\'i Fern\'andez\inst{14}
    \and S. Campana\inst{7}
    \and A. A. Chrimes\inst{1,15}
    \and P. D'Avanzo\inst{7}
    \and V. D'Elia\inst{16}
    \and M. Della Valle\inst{17}
    \and M. De Pasquale\inst{18}
    \and J.~P.~U. Fynbo\inst{2,3} 
    \and N. Gaspari\inst{1}
    \and B. P. Gompertz\inst{19}
    \and D. H. Hartmann\inst{20}
    \and J. Hjorth\inst{5}
    \and P. Jakobsson\inst{21}
    \and E. Palazzi\inst{22}
    \and E. Pian\inst{22}
    \and G. Pugliese \inst{23}
    \and M.~E. Ravasio\inst{1,7}
    \and A. Rossi \inst{22}
    \and A. Saccardi\inst{11} 
    \and P. Schady\inst{24} 
    \and B.~Schneider\inst{25}
    \and J. Sollerman\inst{26}
    \and R.~L.~C. Starling\inst{13}
    \and C. C. Th\"one\inst{27}
    \and A. J. van der Horst\inst{28} 
    \and S. D. Vergani\inst{11,7}
    \and D. Watson\inst{2,3}
    \and K. Wiersema\inst{29}
    \and D. Xu\inst{30}
    \and T. Zafar\inst{31}
    \and S.~Y. Zheng\inst{2,3}
          }

\institute{
    Department of Astrophysics/IMAPP, Radboud University, 6525 AJ Nijmegen, The Netherlands. \\ \email{d.malesani@astro.ru.nl} 
    \and Cosmic Dawn Center (DAWN), Denmark. 
    \and Niels Bohr Institute, University of Copenhagen, Jagtvej 128, 2200 Copenhagen N, Denmark. 
    \and Department of Physics, University of Warwick, CV4 7AL Coventry, United Kingdom 
    \and DARK, Niels Bohr Institute, University of Copenhagen Jagtvej 128, 2200 Copenhagen N, Denmark. 
    \and Artemis, Universit\'e C\^ote d'Azur, Observatoire de la C\^ote d'Azur, CNRS, F-06304 Nice, France. 
    \and INAF, Osservatorio Astronomico di Brera, via E.~Bianchi 46, I-23807 Merate (LC), Italy. 
    \and INFN - Sezione di Milano-Bicocca, piazza della Scienza 3, I-20146 Milano (MI), Italy 
    \and Hessian Research Cluster ELEMENTS, Giersch Science Center, Max-von-Laue-Strasse 12, Goethe University Frankfurt, Campus Riedberg, 60438 Frankfurt am Main, Germany. 
    \and Astrophysics Research Institute, Liverpool John Moores University, 146 Brownlow Hill, Liverpool L3 5RF, UK. 
    \and GEPI, Observatoire de Paris, Universit\'e PSL, CNRS, 5 Place Jules Janssen, 92190 Meudon, France. 
    \and INAF, Istituto di Astrofisica Spaziale e Fisica cosmica, Via Alfonso Corti 12, I-20133 Milano (MI), Italy. 
    \and School of Physics and Astronomy, University of Leicester, University Road, LE1 7RH Leicester, United Kingdom. 
    \and Instituto de Astrof\'isica de Andaluc\'ia - CSIC, Glorieta de la Astronom\'ia s/n, 18008 Granada, Spain. 
    \and European Space Agency (ESA), European Space Research and Technology Centre (ESTEC), Keplerlaan 1, 2201 AZ Noordwijk, the Netherlands. 
    \and ASI - Italian Space Agency, Space Science Data Centre, Via del Politecnico snc, 00133 Rome, Italy. 
    \and INAF, Osservatorio Astronomico di Capodimonte, Salita Moiariello 16, 80131, Napoli, Italy. 
    \and Department of Mathematics, Informatics, Physics and Earth Sciences, University of Messina, via F. D'Alcontres 31, Papardo, Messina, 98166 Italy. 
    \and School of Physics and Astronomy \& Institute for Gravitational Wave Astronomy, University of Birmingham, Birmingham B15 2TT, United Kingdom. 
    \and Department of Physics and Astronomy, Kinard Lab of Physics, Clemson University, Clemson, SC 29634, USA. 
    \and Centre for Astrophysics and Cosmology, Science Institute, University of Iceland, Dunhagi 5, 107 Reykjavík, Iceland. 
    \and INAF–Osservatorio di Astrofisica e Scienza dello Spazio, via Piero Gobetti 93/3, 40024, Bologna, Italy. 
    \and Astronomical Institute Anton Pannekoek, University of Amsterdam, 1090 GE Amsterdam, The Netherlands. 
    \and Department of Physics, University of Bath, Bath BA2 7AY, United Kingdom. 
    \and Kavli Institute for Astrophysics and Space Research, Massachusetts Institute of Technology, 77 Massachusetts Ave, Cambridge, MA 02139, USA. 
    \and The Oskar Klein Centre, Department of Astronomy, Stockholm University, AlbaNova, SE-106 91 Stockholm, Sweden. 
    \and Astronomical Institute of the Czech Academy of Sciences, Fri\v{c}ova 298, 251 65 Ond\v{r}ejov, Czech Republic. 
    \and Department of Physics, George Washington University, 725 21st St NW, Washington, DC, 20052, USA. 
    \and Physics Department, Lancaster University, Lancaster, LA1 4YB, United Kingdom. 
    \and Key Laboratory of Space Astronomy, National Astronomical Observatories, Chinese Academy of Sciences, Beijing, 100101, China. 
    \and School of Mathematical and Physical Sciences, Macquarie University, NSW 2109, Australia
    }            

   \date{Received February 12, 2023; accepted March 16, 2023}

   \titlerunning{The redshift of GRB\,221009A}
   \authorrunning{Malesani et al.}

 
  \abstract
   {The extreme luminosity of gamma-ray bursts (GRBs) makes them powerful beacons, thus effective probes of the distant Universe. The most luminous bursts are typically detected at moderate and high redshift, where the volume for seeing such rare events is maximized and the star-formation activity is greater than at $z=0$. For distant events, not all observations are feasible, such as those at TeV energies.}
   {Here we present a spectroscopic redshift measurement for the exceptional GRB\,221009A, the brightest GRB observed to date, with emission extending well into the TeV regime.}
   {We used the X-shooter spectrograph at the ESO Very Large Telescope (VLT) to obtain simultaneous optical to near-infrared (NIR) spectroscopy of the burst afterglow 0.5 days after the explosion.}
   {The spectra exhibit both absorption and emission lines from material in a host galaxy at \zgrb. Thus, GRB\,221009A was a relatively nearby burst with a luminosity distance of $d_L = 745$ Mpc. Its host galaxy properties (star-formation rate and metallicity) are consistent with those of long GRB hosts at low redshift.
   This redshift measurement yields information on the energy of the burst. The inferred isotropic energy release, $E_{\rm iso} > 5 \times 10^{54}$\,erg, lies at the high end of the distribution, 
   making GRB\,221009A one of the nearest and also most energetic GRBs observed to date.
   We estimate that such a combination (nearby as well as intrinsically bright) occurs between once every few decades and once per millennium.}
   {}


   \maketitle
%

\section{Introduction}

The population of so-called ``long-duration'' gamma-ray bursts (GRBs), which
form via the collapse of massive stars and typically
have prompt phase durations in the range of $\sim 2$ to $1000$ s, constitute the most  
luminous events in the known Universe \citep[e.g.,][]{Kann2007AJ,Racusin2008Nature,Bloom2009ApJ,Frederiks2013ApJ}. They have been detected from a redshift of as small as $z=0.01$ \citep{Galama+98} to $z=8$ \citep{tanvir09,salvaterra09,tanvir18} and possibly beyond $z=9$ \citep{cucchiara11}. Their apparent isotropic-equivalent energies also span a large range. The observed local ($z\lesssim0.2$) burst population is typically of low luminosity, with energies (assuming isotropic emission) of $E_{\rm iso} \sim 10^{48}$--$10^{50}$ erg \citep[e.g.,][]{soderberg04}. 
This reflects the fact that the space density of low-luminosity events is much higher than that of energetic ones.
{In contrast, at high redshift, we are able to sample a much larger volume, where the GRB rate is enhanced compared to the local Universe because of the increase in cosmic star formation. This allows us to detect those comparatively rare GRBs with an isotropic energy release that is  higher by up to six orders of magnitude ($E_{\rm iso} > 10^{54}$ erg; e.g., \citealt{Cenko11}).}
The range of GRB luminosities reflects some observational effects, such as the angle between our line of sight and the axis of the geometrically beamed relativistic jet, but it is also likely that the conditions of the progenitor star at the point of core collapse impact the properties of the subsequent GRB. 

Studies of GRBs in the relatively local Universe are of particular value because their proximity offers diagnostics that are typically not available for the more distant events. 
For example, as the supernovae (SNe) associated with GRBs peak at much fainter magnitudes than the GRB afterglows themselves (especially in the UV), we can study the associated SNe in detail only at $z<0.3$ \citep{2012grb..book..169H,cano17,2014A&A...566A.102S,2015A&A...577A.116D}. Similarly, the lower luminosity distances and better spatial resolution enable studies of their host galaxies and underlying stellar
populations in considerably greater detail \citep[e.g.,][]{Izzo17,2017A&A...602A..85K,2018A&A...615A.136T,2020A&A...633A..68D}. 

\begin{table*}
\begin{center}
\caption{Log of X-shooter spectroscopic observations.}\label{table:spectra}
\begin{tabular}{ccccccc}
\hline\hline
Mid observing time & Time since GRB & Exposure times & Airmass & Seeing & Slit widths   & Slit position angle \\
(UT)               & (days)         & (s)           &         & ($''$) & ($''$)        & (deg)   \\ \hline
2022 Oct 10.0504   & 0.498          & $4\times600$  & 1.8     & 0.85   & 1.0, 0.9, 0.9 & 146.7   \\
2022 Oct 20.0139   & 10.460         & $4\times900$ (VIS), $8\times480$ (NIR) & 1.7     & 0.80   & 1.0, 0.9, 0.9 & 0.0     \\ \hline
\end{tabular}
\tablefoot{The second column shows the time since the \textit{Fermi}/GBM trigger (2022 Oct 9, 13:16:59 UT). The three slit widths refer to the UVB, VIS, and NIR arms, respectively, and the slit position angle is measured counterclockwise from North.}
\end{center}
\end{table*}

Here we consider the case of GRB\,221009A, the brightest GRB observed in over five decades of wide-field-of-view gamma-ray sky monitoring. 
The detection of TeV photons from this GRB
\citep{dzhappuev22,lh23,lh23b} is also unprecedented, both in terms of their energy and flux, making the distance to this event of particular significance, because the mean free path of such photons is limited by their interactions with background light photons \citep{2012MNRAS.422.3189G,2017A&A...603A..34F}.
Our X-shooter spectroscopy of the burst provides a robust redshift measurement of $z=0.151$, consistent with being a cosmological GRB and not a transient originating in the Milky Way (MW) as its location on the sky may suggest. The implied luminosity distance\footnote{Assuming a flat $\Lambda$CDM cosmology with $H_0 = 67.4$~km~s$^{-1}$~Mpc$^{-1}$ and $\Omega_m = 0.315$ \citep{2020A&A...641A...6P}.} of $d_L = 745$\,Mpc places GRB\,221009A amongst the most energetic GRBs ever observed, and makes it by far the closest GRB with $E_{\rm iso} > 10^{54}$ erg.

\section{Observations}
\subsection{High-energy discovery}
GRB\,221009A was detected at 2022 October 9 at 13:16:59 UT by a raft of high-energy missions
including \textit{Fermi}/GBM \citep{veres22,lesage23}, \textit{Fermi}/LAT \citep{bissaldi22}, AGILE/MCAL and GRID \citep{ursi22,piano22,tavani23}, 
INTEGRAL \citep{gotz22,rodi23}, Konus/\textit{Wind} \citep{frederiks22,frederiks23}, INSIGHT/HMXT \citep{tan22,kann23}, STPSat-6/SIRI-2 \citep{mitchell22}, SATech-01/GECAM-C HEBS \citep{liu22}, SRG/ART-XC \citep{2022GCN.32663....1L}, Solar Orbiter/STIX \citep{2022GCN.32661....1X}, 
and GRBalpha \citep{ripa22,ripa23}. However, the event was first reported by a \textit{Swift} detection of the afterglow over 50 minutes later \citep{2022GCN.32632....1D}. The location of the burst within the Galactic plane ($l=52.96^\circ$, $b=4.32^\circ$), combined with its brightness, led to confusion over the nature of the outburst: initially it was suspected to be due to a new Galactic X-ray transient \citep{2022GCN.32632....1D,2022ATel15650....1D}, but its subsequent behavior, together with the \textit{Fermi}/GBM and LAT detections, appeared more like that of an extragalactic GRB \citep{2022GCN.32635....1K,veres22,bissaldi22}.

Despite high foreground extinction (Sect.~\ref{sec:extinction}), an optical afterglow was seen by various telescopes (e.g., \citealt{2022GCN.32632....1D,2022GCN.32634....1L,Fulton2023} and many more).
The counterpart was localized at coordinates (J2000):
$\text{RA} = 19^{\rm h}13^{\rm m}03\fs500792$(2), $\text{dec} = 19^\circ46'24\farcs22891$(7) by the VLBA at 15.2 GHz \citep{2022GCN.32907....1A}.

Detections with several high-energy instruments have also been reported, including GeV emission with
\textit{Fermi}/LAT (potentially up to 400 GeV; \citealt{xia23}), TeV emission extending to 18 TeV from LHAASO \citep{lh23}, and even a suggestion of a possible association with a 250 TeV photon \citep{dzhappuev22}.

\subsection{X-shooter spectroscopy} 

Following the detection of GRB\,221009A, and motivated by the significant ambiguity in its distance (Galactic versus extragalactic), we initiated observations with the European Southern Observatory Very Large Telescope (ESO VLT Unit 3, Melipal) located on Cerro Paranal (Chile) and the X-shooter spectrograph \citep{2011A&A...536A.105V}. These
observations began on 2022 October 10 at 00:49:26 UT (11.54 hr after the \textit{Fermi} trigger). The source brightness at this time, as measured in our acquisition image, was $r^\prime=17.42\pm0.05$ mag (AB, calibrated against nearby stars from the Pan-STARRS catalog). {A summary of the X-shooter spectroscopic observations is presented in Table~\ref{table:spectra}.}

{Because of the limited target visibility, the airmass at the time of the observation was sizeable, but within the working range of the X-shooter atmospheric dispersion corrector (ADC).} Observations were executed using the ABBA nod-on-slit mode, with a nod throw of about $6''$ along the slit. Each single arm spectrum was reduced in ``stare'' mode and using the standard X-shooter pipeline \citep{Goldoni2006,Modigliani2010}, with the extraction window at the position of the GRB afterglow trace, and background windows at both sides of the spectral trace. Then, for each exposure, 
residual sky features were interpolated using the background as reference \citep{Selsing19}. {Flux calibration was performed in two steps, first by applying the response function determined via observations of spectrophotometric standard stars, and then renormalizing the flux-calibrated spectra to the available photometry} using the $g$, $i,$ and $H$ bands for the UVB, VIS, and NIR arms, respectively, interpolated at the mean time of the spectrum. Magnitudes were computed from images secured with the X-shooter acquisition camera or taken from a broader photometric set (\citealt{2022GCN.32652....1B}; de Ugarte Postigo et al., in prep.). Finally, we applied a telluric correction built using a line-by-line radiative transfer model (LBLRTM; \citealt{Clough1992}) and the atmospheric properties, such as humidity, temperature, pressure, and zenith angle, which are stored in the header of each exposure.

\begin{figure*}
   \centering
   \includegraphics[width=\textwidth]{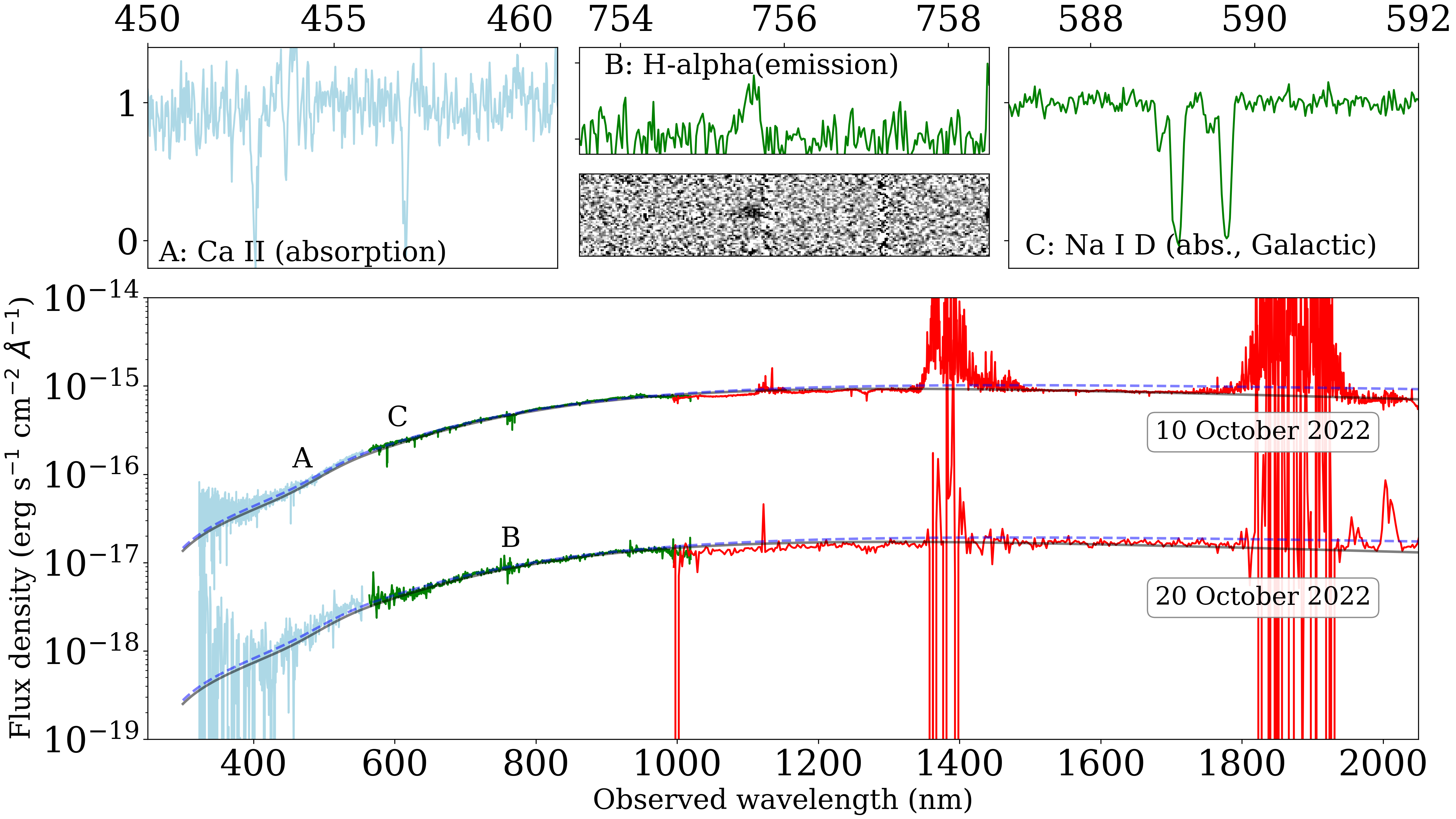}
      \caption{Our X-shooter spectra of the afterglow of GRB\,221009A on 2022 October 10 and 20. For clarity, the spectra shown in the bottom panel have been rebinned by a factor of 10 (first epoch) and 40 (second epoch). The UVB, VIS, and NIR arms are color coded in cyan, green, and red, respectively. The blue dashed lines show a power law with a spectral index of $\beta = 0.8$ ($F_\nu \propto \nu^{-\beta}$), extinguished by a foreground extinction of $A_V=4.177$ mag, assuming the MW extinction law from \citet{CCM89}. The solid black lines show a bluer $\beta=0.4$ spectrum with $A_V = 4.7$ mag, and provide a slightly better description of the data. The upper panels show zoomed-in regions around the \ion{Ca}{ii} absorption and H$\alpha$ emission at $z=0.151$ (the afterglow continuum was digitally subtracted from the 2D plot) as well as the \ion{Na}{i} D absorption complex  due to dust in the MW.}
      \label{XS_main}
   \end{figure*}

The observations reveal a very bright trace in the red and infrared that is strongly attenuated toward the blue end by the high Galactic extinction.
Figure~\ref{XS_main} shows the overall shape of the spectrum and the inner panels zoom in on selected features.
We subsequently obtained further X-shooter observations to follow the afterglow evolution. These are discussed in detail by de Ugarte Postigo et al. (in prep.). Among the late spectra, here we only present the $8\times 480$~s spectrum, which has a mid time of 2022 October 20 00:20:01 UT and provides the best detection of the emission features (Fig.~\ref{XS_main} and Sect.~\ref{sec:host}).

The results reported in this paper supersede our preliminary analysis \citep{2022GCN.32648....1D,2022GCN.32765....1I}. Our spectroscopic measurement was subsequently confirmed by \citet{2022GCN.32686....1C}.

\section{Results}

\subsection{Absorption line system and redshift}

In addition to several Galactic absorption features at $z = 0$, we identify a series of absorption lines including Ca\,\textsc{ii} (4530, 4570\,\AA), Ca\,\textsc{i} (4867\,\AA), and Na\,\textsc{i} (6782, 6789\,\AA) at redshift $z \approx 0.151$. Unfortunately, at this low redshift, most of the strong metal lines are too far into the UV to be detected from the ground, and so we cannot carry out any in-depth analysis of the GRB host ISM.

\begin{figure}
   \centering
   \includegraphics[width=\hsize]{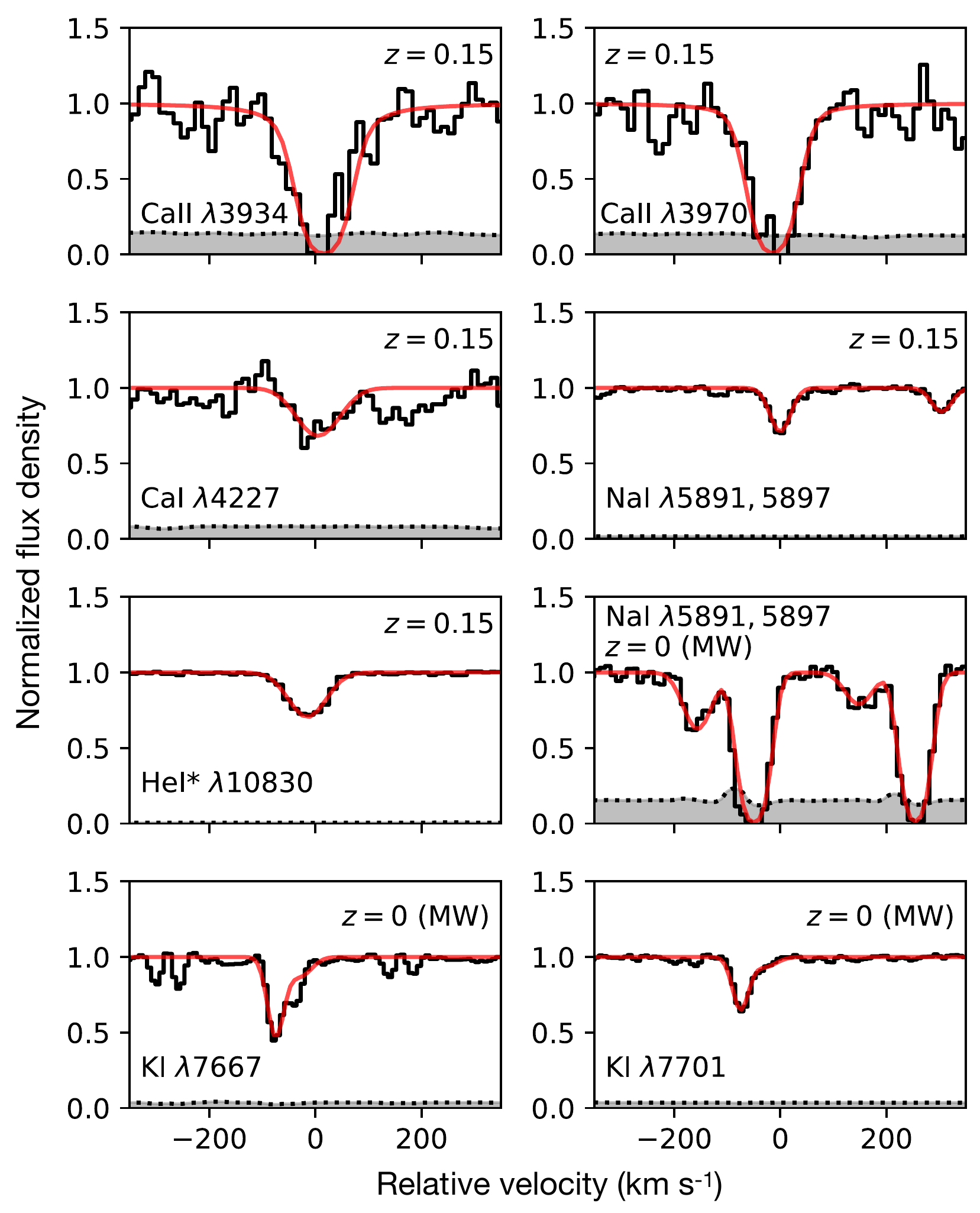}
      \caption{Zoomed-in view of selected absorption features in velocity space. The gray-shaded areas show the error spectrum, while the red lines indicate the best-fit model. The \ion{K}{i} 7667~\AA{} line is affected by telluric absorption and was not used for the column density determination.}\label{fig:abslines}
   \end{figure}

We first analyzed the absorption features imprinted from the host-galaxy ISM on the afterglow spectrum. To this end, we used the \texttt{Python} package \texttt{VoigtFit} \citep{Krogager2018}. This code creates a convolved spectrum based on the observed spectral resolution of the X-shooter spectra and simultaneously fits for the redshift $z_{\rm GRB}$, broadening parameter $b$, and column density of each line complex. We tie the redshift to be identical across the five detected transitions in the optical (see Table \ref{table:abslines}). The best-fit redshift is $z_{\rm GRB} = 0.15095\pm 0.00005$. We also find that a single broadening parameter $b=18.5\pm 5.0$\,km\,s$^{-1}$ is able to reproduce the observed line profiles (see Fig.~\ref{fig:abslines}), with total column densities of $\log (N_\ion{Ca}{ii}/\text{cm}^{-2}) > 15.5$, $\log (N_\ion{Ca}{i}/\text{cm}^{-2}) = 12.19\pm 0.08$, and $\log (N_\ion{Na}{i}/\text{cm}^{-2}) = 12.21\pm 0.01$. We note that the \ion{Ca}{ii} transitions are both saturated, and so the inferred column densities are quoted as $3\sigma$ lower limits.

{We further note the presence of a well-detected absorption feature in the NIR arm observed at 12,468~\AA, which we cannot associate with any telluric or Galactic transition. However, it matches the \ion{He}{i}{}$^*$ 10,830~\AA{} (unresolved) triplet at $z = 0.151$. The line profile reveals that the redshift and Doppler broadening of this particular feature are not consistent with the main ones described above, likely suggesting a different physical origin. From the best-fit Voigt profile, we find $z_{\rm HeI*} = 0.15089\pm 0.00003$, $b=40.9\pm 1.3$\,km~s$^{-1}$, and a column density with $\log (N_\ion{He}{i*}/\text{cm}^{-2}) = 12.29\pm 0.01$.
This feature was also identified by \citet{Fynbo_140506A} in the peculiar afterglow of GRB\,150406A at $z = 0.889$, and is also visible in the absorption spectrum of GRB\,190114C at $z = 0.425$ \citep{deUgartePostigo_190114C}.}

\begin{table}
\caption{Host absorption lines identified in the first X-shooter spectrum, with their measured equivalent widths (observer frame) and column densities.}\label{table:abslines}
\begin{tabular}{llll}
\hline\hline
Observed $\lambda$ & Feature  & EW    & $\log (N/\mathrm{cm}^{-2})$             \\ 
(\AA)              &          & (\AA) & \\
\hline  
4530.09  & \ion{Ca}{ii} 3934.78  & 1.58$\pm$0.1   & \multirow{2}{*}{$>15.5$} \\ 
4570.44  & \ion{Ca}{ii} 3969.59  & 1.35$\pm$0.09  &                          \\ 
4867.40   & \ion{Ca}{i}  4227.92  & 0.36$\pm$0.06  & $12.19\pm 0.08$          \\
6782.87  & \ion{Na}{i}  5891.58  & 0.29$\pm$0.01  & \multirow{2}{*}{$12.21\pm 0.01$} \\ 
6789.71  & \ion{Na}{i}  5897.56  & 0.15$\pm$0.01  &                          \\
12470.76 & \ion{He}{i}$^*$ 10830    & 1.025$\pm$0.015 & $12.29\pm 0.01$\tablefootmark{a} \\
\hline
\end{tabular}
\tablefoot{\tablefoottext{a}{Best-fit column density with $z_{\rm \ion{He}{i}^*} = 0.15089\pm 0.00003$ and $b=40.9\pm 1.3$\,km s$^{-1}$.}}
\end{table}

\subsection{Extinction}\label{sec:extinction}

The line of sight to GRB\,221009A goes through significant foreground extinction from dust in the MW and potentially also within its host galaxy. The large foreground extinction is highlighted by strong absorption features seen at $z = 0$, such as \ion{K}{i} and the strongly saturated \ion{Na}{i} D (Fig.~\ref{XS_main}). In Fig.~\ref{XS_main}, we overplot the X-shooter spectra with a single power law with index $\beta = 0.8$ ($F_\nu \propto \nu^{-\beta}$; the same as observed in the X-ray afterglow, e.g., \citealt{Williams2023}) with a foreground extinction of $A_V = 4.177$ mag \citep{SF11}. This provides a reasonable description of the observations, but substantially over-predicts the measured flux in the red. This suggests the presence of additional extinction,
either in the MW (a small underestimate, particularly toward the plane of the MW, is not unlikely)
or due to additional dust inside the host galaxy. Given the low redshift of the event and the diversity of the extinction laws \citep{pei92}, disentangling host from foreground extinction is nontrivial, but the overall results are relatively insensitive to the dust location. 
{A better match to the data can be obtained with a larger total $A_V$. However, in this case the intrinsic spectral slope must be substantially bluer, matching the preferred results of \cite{kann23} and \cite{levan2023} well. 
A Monte Carlo Markov Chain sampling of the data using a single power-law source with extinction described by a \cite{fitzpatrick1999} extinction law returns $A_V = 4.9$ mag and $R_V = 3.225$, with negligible host-galaxy extinction, and a power law with a spectral index of $\beta = 0.21$.
This would be consistent with the presence of a cooling break lying between the X-ray and optical/IR band.}

We also note that there is a significant revision in the foreground extinction between the map of \cite{schlegel98} and the recalibration of \cite{SF11}, which we have adopted, with the former value of $A_V = 4.8$ mag overall being consistent with our higher extinction scenario. Further insight into the dust location can be provided by the detection of \ion{Na}{i} D at $z = 0.151$, which is often considered a dust tracer. Following  \cite{poznanski12} and using the EWs from Table \ref{table:abslines}, we derive $A_V = 0.14^{+0.15}_{-0.11}$ mag, suggesting that indeed a (modest) amount of dust is present in the GRB host galaxy along its line of sight, but that extra dust is also present in the MW, in excess of the \citet{SF11} estimate.

\subsection{Host galaxy}\label{sec:host}

\begin{figure}
    \includegraphics[width=\columnwidth]{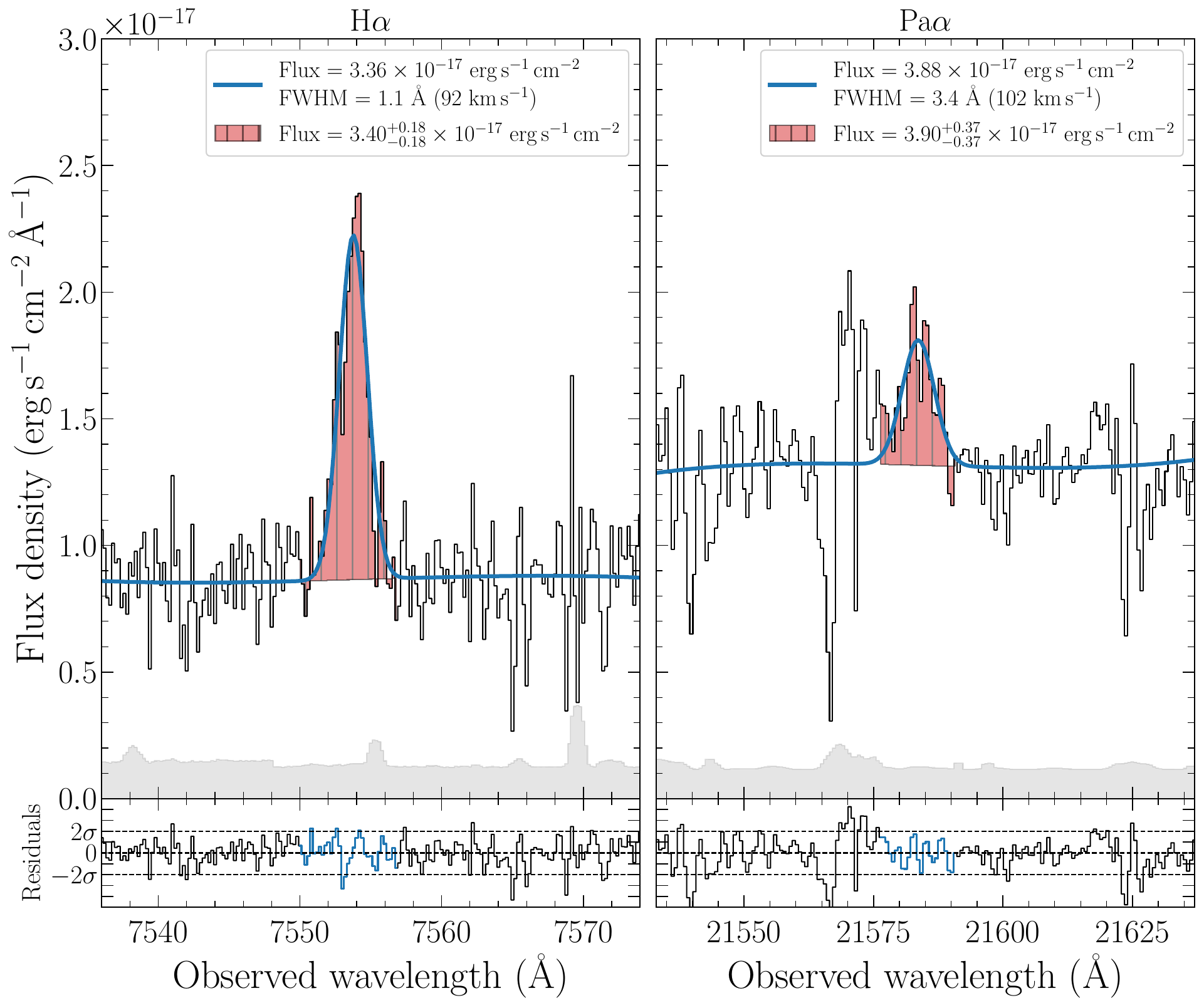}
    \caption{Single-Gaussian fit to the \Ha (left) and \Pa (right) emission lines. A sky line is responsible for the large residuals blueward of \Pa.
    {} Top panel: The blue line is the result of the fit of the line and the continuum whose fit parameters are reported in the legend (the FWHM is measured in the  observer frame).
    The red shaded area represents the direct integration of the observed flux, whose value is reported in the legend; this is the value used in the calculations throughout the paper.
    The gray shaded area represents the error spectrum.
    {Bottom panel:} Normalised residuals. The blue points represent the region where the line is fit.
    }\label{fig:Ha_Pa_fit}
\end{figure}

In addition to absorption lines, our X-shooter spectroscopy shows two emission lines from the underlying host galaxy.
We detect \Ha (in the visible) and \Pa (in the NIR), while we cannot satisfactorily recover the bluer \OIIIb, \OIIIa, \Hb, or \OII, presumably due to the higher foreground extinction.
A small spatial offset is observed between the afterglow trace and the \Ha emission line, which is consistent with the extension seen in {\em Hubble} Space Telescope imaging of the field (\citealt{2022GCN.32921....1L,levan2023,Fulton2023,Shrestha2023}; see also \citealt{blanchard24}). This also implies that the measured fluxes are lower limits, as there is likely emission from regions not covered by the instrument slit. The slit was 0.9\arcsec{} (1.1\arcsec) wide in the optical (NIR), and was oriented along the N--S direction, only partially
covering the extended host galaxy. Line-flux ratios are however robust against slit losses.
The flux of the detected lines was determined by fitting the continuum around each line and integrating the continuum-subtracted flux over the region of the line.
We additionally fitted each detected line with a Gaussian profile, checking the consistency between the two values. The fits are shown in Fig.~\ref{fig:Ha_Pa_fit}.
We also computed an upper limit for \NIIb as the flux of a Gaussian line with the same velocity width as \Ha and an amplitude equal to the noise estimated over the region where the line should lie.

The fluxes were then corrected for Galactic dust extinction using $A_V = 4.177$ mag and the MW extinction curve of \citet{pei92}.
We determined the host galaxy $A_V$ comparing the observed and theoretical ratio of the \Ha and \Pa lines ($F(\Ha)/F(\Pa) = 8.56$) in the case B recombination scenario, with an electron temperature of 10,000~K and a density of $10^4$~cm$^{-3}$ \citep{osterbrock06}.
This method yields \Avhost~mag, where the uncertainties are computed with Monte Carlo error propagation. As with the afterglow SED, it is not possible to disentangle whether this extra extinction is due to dust in the MW or in the GRB host, but this has little effect on our estimate given the low redshift of the object.
All the fluxes and their corrected values are shown in Table~\ref{tab:em_line_fluxes}.

Using the dust-corrected \Ha flux and the relation of {\citet{Kennicutt98}}, we infer a star-formation rate (SFR) of \SFRhost \Msunyr, scaled to the initial mass function of \citet{chabrier03}, setting \zem and a standard \citet{2020A&A...641A...6P} cosmology.
Due to slit losses, this value should be considered as a lower limit.

Using the upper limit on the \NIIb line, we can infer an upper limit on the metallicity of the host galaxy using the strong line calibrations of \citet{maiolino08}, and taking $\rm 12+\log(O/H)_{\odot} = 8.69$ \citep{Asplund09}.
We find \logOHhost which is subsolar and is comparable to the typical metallicities found using the same method in long GRB hosts \citep{kruehler15,japelj16,palmerio19}.

\begin{table*}
\begin{center}
\caption{Fluxes of the host emission lines measured in the X-shooter spectrum from October 20. The last two columns report the line fluxes corrected for the MW extinction ($A_V = 4.177$~mag) and MW + host, respectively.}
\begin{tabular}{ccccc}
\hline\hline
Line & Observed wavelength & Observed flux & Corrected flux (MW) & Corrected flux (MW + host)\\
& (\AA) & ($10^{-17}$\,\ergscm) & ($10^{-17}$\,\ergscm) & ($10^{-17}$\,\ergscm) \\ \hline
\Ha & $7553.80$ & $3.4^{+0.2}_{-0.2}$ & $39.0^{+2.0}_{-2.0}$ & $51.0^{+5.7}_{-5.6}$\\

\Pa & $21582.75$ & $3.9^{+0.4}_{-0.4}$ & $5.7^{+0.5}_{-0.5}$ & $6.0^{+0.7}_{-0.7}$ \\
\NIIb & --- & $< 0.4$ & $< 4.0$ & $< 5.3$ \\ \hline
\end{tabular}
\end{center}\label{tab:em_line_fluxes}
\end{table*}

\section{Discussion}

\subsection{Robustness of redshift}
Given the challenges of explaining VHE emission from $z=0.151$, it is
relevant to consider whether or not there are other, lower redshift 
possibilities able to explain the burst. The presence of narrow absorption lines implies that the light from the GRB is passing through cool, low-velocity gas moving at a recessional velocity of $\sim 45{,}000$ km s$^{-1}$. This is well beyond any plausible peculiar velocities for more local galaxies, or possible velocities of individual gas clouds within them. Although outflows (e.g., from SNe) can achieve high velocities, most do not reach 45,000 km s$^{-1}$, and in this case we would then expect to observe broad lines. Furthermore, the presence of narrow emission lines at $z=0.151$ demonstrates that a star-forming galaxy at this redshift underlies the GRB position. The detection of a SN akin to SN\,1998bw \citep{blanchard24} further confirms the redshift value reported in this work.

{We estimate the chance alignment probability $P_{\rm chance}$ between the afterglow and the galaxy using the formula presented by \citet{2002AJ....123.1111B}. Using HST imaging, \citet{levan2023} report a $0.25\arcsec$ angular separation between the host centroid and the afterglow, and a host effective radius of $\approx 1\arcsec$. In this case, it is the latter quantity that should be used to compute $P_{\rm chance}$. Using the host magnitude $\mbox{F625W} = 24.88$~AB, and correcting for the foreground extinction, we find $P_{\rm chance} = 1$--$1.5\times10^{-3}$, where the uncertainty comes from the unknown proportion of extinction within the MW. Independent of this, the chance alignment is very small, further solidifying the association with the galaxy and confirming the robustness of the GRB redshift identification.}

Formally, we note that the detection of narrow absorption lines only places a lower limit on the burst redshift of
$z = 0.151$. The redshift is such that we cannot detect individual variable fine-structure lines pumped by the
UV emission from the GRB \citep[e.g.,][]{vreeswijk04}. It is plausible that the burst could be at even higher redshift. A higher redshift would ameliorate rate constraints, and could even be lensed by the $z=0.151$ system (although such a scenario is very unlikely). However, placing the burst at an even higher redshift would only exacerbate the challenges of VHE emission and extreme prompt isotropic energy release. The lack of any detected features at redshifts of higher than $z=0.151$ (including across wavelength ranges with high S/N) also argues against $z > 0.151$.

\subsection{Energetics and implications}
GRB\,221009A is by far the 
brightest GRB ever observed \citep{Burns2023}. Measurements of both the peak flux and fluence are underestimated because it was sufficiently bright to saturate many of the 
detectors, but even allowing for this, GRB\,221009A was a factor of 40 more fluent than the next-brightest well-studied burst at the time%
\footnote{During the review process of this paper, another very bright event was observed, which has now become the second brightest detected GRB; \citep{2023GCN.33414....B}.},
GRB\,130427A {\citep{2014Sci...343...42A,2014Sci...343...51P,maselli14}}.

Estimating the precise isotropic energy release of GRB\,221009A is
nontrivial  because of the
difficulty in correcting for saturation effects. However, estimates of the fluence lie in the range
$\approx(\mbox{1--9})\times10^{-2}$ erg~cm$^{-2}$ \citep{gotz22,2022GCN.32642....1L,frederiks22,2022GCN.32762....1K} and the
burst isotropic energy release in the 0.1~keV--100~MeV ``ultrabolometric'' band \citep{2023MNRAS.tmp..186F} is at least
$E_{\rm iso} =  5.9 \times 10^{54}$ erg \citep{2022GCN.32762....1K}.  In addition to being one of the closest, this makes GRB\,221009A one of the most intrinsically energetic GRBs ever observed. If a more detailed saturation correction analysis yields a true fluence of $\gtrsim10^{-1}$ erg cm$^{-2}$, GRB\,221009A could well also be the most energetic GRB ever observed, at least within the sample of GRBs with known redshift.

\subsection{Rates}
\begin{figure}
    \includegraphics[width=\columnwidth]{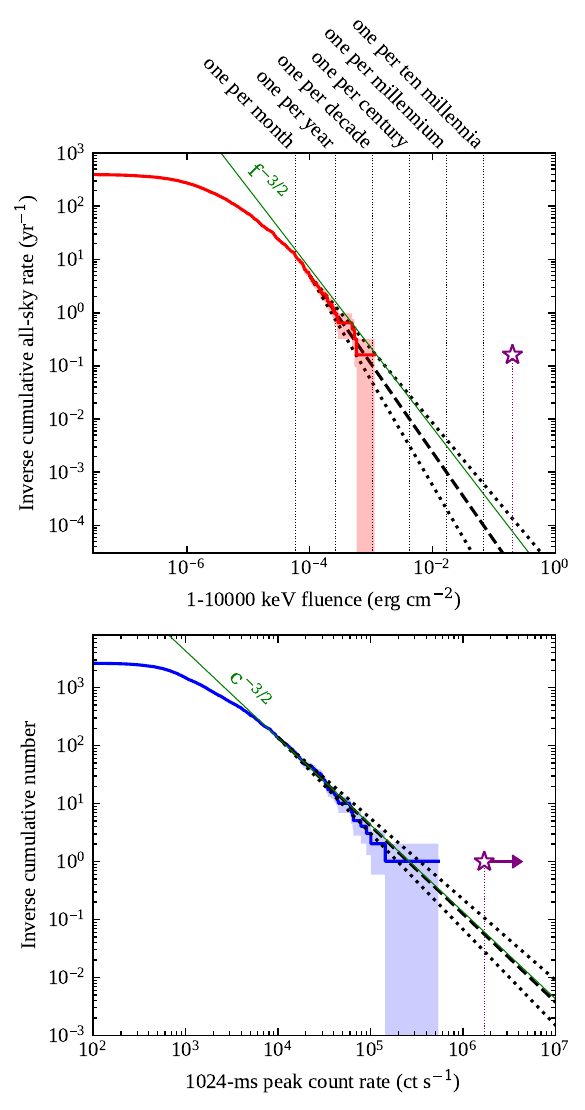}
    \caption{Differential and integrated GRB rates. {Top panel:} {All-sky cumulative rate} of long GRBs (i.e.,\ those with $T_{90}>2$\,s; thick solid line) observed by \textit{Fermi}/GBM up to 2018 June above a given 1--10\,000 keV fluence, excluding GRB\,221009A. {The estimated fluence of the latter, $\sim 0.2$~erg\,cm$^{-2}$, based on \textit{Fermi}/GBM \citep{2022GCN.32642....1L,Burns2023}, Konus/\textit{Wind} \citep{frederiks23}, INSIGHT-HXMT, and GECAM-C \citep{An2023},} is shown by the purple star. The shaded area comprises the distributions obtained by varying the observed number by $\sqrt{N}$, where $N$ is the number of bursts in each bin, which approximates the Poisson error. The black dashed line shows the best-fit power-law model $N(>f) = N_0(f/f_0)^{-a}$, where $f_0 = 10^{-4}$ erg cm$^{-2}$ and $N_0$ is the observed rate above $f_0$. The black dotted lines encompass the 68\% uncertainty on the power-law index $a$ from such a fit. The thin green line shows the theoretically expected $-3/2$ scaling for homogeneously distributed sources in Euclidean space {although there is no strong expectation that GRBs should follow this distribution (see text)}. {The vertical dotted lines mark the fluxes that correspond to the reported rates according to the best-fit power-law model.} {Bottom panel:} Similar to the top panel, but showing the cumulative number of events above a given peak background-subtracted count rate (summed over all channels and all NaI detectors, measured with 1024 ms binning, {and {not} corrected for saturation}). In this case, $f_0=10^4\,\mathrm{ct\,s^{-1}}$, {and the peak count rate of GRB\,221009A is a lower limit}.}
    \label{fig:logN_logS}
\end{figure}

The combination of proximity and
brightness or energy release raises the question of just how unusual GRB\,221009A is. To place this GRB in context, we compared it with long GRB events (i.e., those with $T_{90}>2\,\mathrm{s}$) previously observed by \textit{Fermi}/GBM. In general, it is relevant to consider both the 
peak flux and the fluence. While the peak flux is normally more relevant for triggering, in the case of GRB\,221009A it is also likely 
more 
{difficult to correct for}
saturation compared to fluence.
We thus considered both the {``bolometric'' 1--10\,000 keV} band fluence\footnote{We used the information available in the online \textit{Fermi}/GBM catalog, accessible at \url{https://heasarc.gsfc.nasa.gov/W3Browse/fermi/fermigbrst.html}, {focusing on bursts detected up to November 26 2018, which were published in the fourth \textit{Fermi}/GBM catalog \citep{vonKienlin2020}. For bursts with available spectral information, we used the best-fitting spectral model (as reported in the \texttt{flnc\_best\_fitting\_model} column) with the reported best-fit parameters to compute the bolometric fluence. For the remaining $\sim 5.8\%$ of the bursts, we used the 10--1000 keV value reported in the \texttt{fluence} column, which is a good approximation of the bolometric fluence, assuming the peak of the $\nu F_\nu$ spectrum is below 1000 keV}.} and the peak count rates, calculated with 1024 ms binning, summed over all NaI detectors and all channels. To construct the latter, we downloaded all \texttt{trigdata} files related to long GRBs from the \textit{Fermi}/GBM trigger catalog\footnote{\url{https://heasarc.gsfc.nasa.gov/W3Browse/fermi/fermigtrig.html}} and used the count-rate data therein. In 98\% of cases, peak count rates were reported with a binning of 1024 ms or finer (in which case we downsampled to 1024 ms by summing neighboring bins near the peak). We assumed the remaining 2\% of the events to be sampled from the same distribution. From these, we constructed the corresponding inverse cumulative distributions, which are shown in Fig.~\ref{fig:logN_logS}. In order to convert the number of observed events into an all-sky rate, we divided the number by the GBM mission duration ($T_\mathrm{GBM}=10$~yr, {as we used only bursts whose spectral information is part of the fourth \textit{Fermi}/GBM catalog}; \citealt{vonKienlin2020}), and by the effective GBM duty cycle $\eta=0.598$, determined by its limited field of view and downtime, itself attributable to the need to turn off the detectors during the transit through the South Atlantic anomaly \citep{burns2016}. In order to extrapolate the observed inverse cumulative distribution to the 
fluence (or count rate) of GRB\,221009A, we identified a 
fluence (or count rate) $f_0$ above which the distribution resembles a single power law and fitted a simple power-law model with index $a$ to the data above that point {(excluding GRB\,221009A itself)}. The fit is performed in a Bayesian fashion, as follows. The assumed probability of observing a given 
fluence (or count rate) $f_i$ is $p(f_i\,|\,a)=(a/f_0)(f/f_0)^{-a-1}$, and therefore the posterior probability on $a$ given the observed 
fluences (or count rates) is
\begin{equation}
    p(a\,|\,\vec{d})\propto\pi(a)\prod_{i=1}^{N_0}\frac{a}{f_0}\left(\frac{f_i}{f_0}\right)^{-a-1},
\end{equation}
where $\vec{d}=\lbrace f_i \rbrace_{i=1}^{N_0}$ is our data and $\pi(a)$ is the prior on $a$, which we take as uniform in $a$, $\pi(a)\propto 1$. We set $f_0=10^{-4}\,\mathrm{erg\,cm^{-2}}$ for 
fluences and $f_0=10^4\,\mathrm{ct\,s^{-1}}$ for count rates. The posterior probability density distributions on $a$ are nearly Gaussian for both the fluence and count rate cases, and yield $a\sim 1.66_{-0.28}^{+0.31}$ for fluences
and $a\sim 1.52\pm0.13$ for count rates (median and symmetric 68\% credible interval). {While the slope in both cases is close to $a=3/2$, which is expected for homogeneously distributed sources in Euclidean space, this is not, in this case, indicating such a distribution, but is caused by the increase in the GRB rate with redshift \citep[$\propto (1+z)^{3.2}$;][]{Ghirlanda2022}, which, in this calculation, cancels out the effect of the expansion of the Universe.} Given the power-law scaling, the expected all-sky rate of events above a fluence (or count rate) $f^\star$ is simply
\begin{equation}
    \bar R_\mathrm{obs}(f\geq f^\star)=\frac{N_0}{T_\mathrm{GBM}\eta}\left(\frac{f^\star}{f_0}\right)^{-a}.
\end{equation}
Figure \ref{fig:logN_logS} shows the {resulting expected rate extrapolation, which is compared to the GRB\,221009A bolometric fluence $f^\star\sim 0.2\,\mathrm{erg\,cm^{-2}}$ \citep{frederiks23,Burns2023} and to its peak count rate of $f^\star=1.7\times 10^{6}\,\mathrm{ct\,s^{-1}}$ obtained as explained above. The latter is a lower limit because of the saturation in the NaI detectors.}

Taking the values corresponding to the 90$^\mathrm{th}$ percentiles {of the posterior distributions, we estimate a rate of} less than 1 event in 
{4210} years (fluences) or {70}
years (peak count rate) as bright as or brighter than GRB\,221009A. {The former value is} 
{much lower than that}
found by \cite{2022GCN.32793....1A}, {owing to the fact that we employ a bolometric, saturation-corrected estimate of the brightness of this burst}.

Given these expected rates, the probability of observing at least one event {with \textit{Fermi}/GBM}, assuming Poisson statistics, is given by
\begin{equation}
\begin{split}
    p(N(f\geq f^\star)>0\,|\,\vec{d}) = 1-p(N(f\geq f^\star)=0\,|\,\vec{d}) =\\
    = 1-\int \exp(-\bar R_\mathrm{obs}\eta T_\mathrm{GBM}) p(a\,|\,\vec{d})\,\mathrm{d}a, 
\end{split}
\end{equation}
which yields  
$p(N(f\geq f^\star)>0\,|\,\vec{d})\sim 6.8\times 10^{-4}$ using the fluences, 
and $p(N(f\geq f^\star)>0\,|\,\vec{d})\lesssim 0.064$ using the count rates. Again, the {latter} must be taken as an upper limit due to saturation. These values 
confirm that GRB\,221009A is a very rare event, {in agreement with \citet{Burns2023}}. 

\subsection{Estimate based on a  GRB population model}

We also provide an estimate of the rate of GRB\,221009A-like events based on the state-of-the-art model for the population of long GRBs described in \cite{Ghirlanda2022}. This model predicts the intrinsic properties of the population of long GRBs, such as their luminosity function and cosmic rate, as obtained by reproducing the observed distributions of the properties of bursts detected by \textit{Fermi}/GBM and CGRO/BATSE, and the distributions of rest-frame properties of a flux-limited sample of bright GRBs detected by {\it Swift}/BAT \citep{Salvaterra2012}. By sampling the posterior distribution of the model population parameters (Table~1 in \citealt{Ghirlanda2022}), we estimated the probability of occurrence of an event with an isotropic-equivalent luminosity $L\ge L_{\star}$ within a redshift of $z\le z_{\star}$, {where $L_{\star}=10^{54}$ erg s$^{-1}$ \citep{Burns2023} and $z_{\star}=0.15$ correspond to the values of GRB\,221009A.} With these values, we sampled the population model posterior probability {and derived a mean rate of 1 every 18\,400 yr (or 1 every 3150 yr as 90$^{\rm th}$ percentile estimate).}

In the entirety of 55 years of GRB observations, no event has been reported that even comes close to GRB\,221009A in terms of fluence \citep{Burns2023}. {Prior to} GRB\,221009A, GRB\,130427A was very likely the most fluent GRB of the combined CGRO/BATSE, Konus/\textit{Wind}, and \textit{Fermi}/GBM era beginning in 1991, which is well documented and readily publicly available \citep{1999ApJS..122..465P,2006ApJS..166..298K,2011A&A...530A..21N,2013ApJS..208...21G,Tsvetkova2017ApJ,Tsvetkova2021ApJ,2019ApJ...878...52A,2020ApJ...893...46V}. {More recently, \textit{Fermi} detected another high-fluence event, namely GRB\,230307A with a preliminary estimate of the fluence of $\sim 3\times 10^{-3}$ erg cm$^{-2}$ \citep{dalessi23}, comparable to that of GRB\,130427A and also a very rare (once per few decades) event \citep{burns23b,levan23b}. 
}

The situation for the first $\sim25$ years of GRB studies is less transparent, but there are several events with fluences similar to that of GRB\,130427A, namely the famous GRB\,830801B \citep{1986SvAL...12..315K,1987PAZh...13.1055K,2001ApJ...563L.123S}, the significantly less well-documented GRB\,840304 \citep{1984BAAS...16Q1016K,1985BAAS...17..850I}, and two events detected by the Pioneer Venus Orbiter (PVO), GRBs\,920212 and 900808 (the latter mentioned in \citealt{1996ApJ...460..964F} as the brightest GRB in terms of peak flux detected by PVO). However, none of these events had a fluence in excess of $10^{-2}$ erg cm$^{-2}$, and so GRB\,221009A is potentially more than ten times as fluent as any other GRB ever detected \citep{Burns2023}.

\subsection{VHE emission} 

One of the most striking features of the detection
of GRB\,221009A is the identification of very high-energy
emission. In particular, the LHAASO water Cherenkov observatory reported the detection of more than 5000 TeV photons associated with the burst 
\citep{huang22,lh23,lh23b}, with a highest energy photon at 18 TeV. The Carpet-2 experiment detected 
a possible 250 TeV photon \citep{dzhappuev22}. 

Knowledge of the redshift is critical for the interpretation of these observations, because TeV photons scatter off the extragalactic background light (EBL) photons or even ---at the very highest energies--- off the cosmic microwave background (CMB) photons. The EBL is the result of photons from essentially every photon emitter in the Universe, and is therefore, in addition to the CMB, of important cosmological utility. 
These photon scatterings result in progressively higher opacity for photons of increasing energy.
The detection of high-energy TeV photons is therefore a valuable route to probing models for the EBL.

Indeed, the detection of such emission from GRB\,221009A at $z=0.151$ is particularly intriguing.
For example, by reconstructing the expected intrinsic emission from blazars, 
\citet{dominguez13} estimated the effective horizon as a function
of photon energy: at 18 TeV this horizon (corresponding to an 
optical depth of 1, or an attenuation factor of $0.63$) lies at $z<0.01$, while at $z=0.15$, $\tau \gg 3$ (attenuation $>0.95$). {In practice, the situation is less severe, because the uncertainty associated with the highest-energy photons is significant (formally \cite{lh23b} find $E=17.8^{+7.4}_{-5.1}$ TeV) and because alternative, nonphotonic (e.g., muon) origins cannot be  robustly ruled out. However, there remains some tension that may be suggestive of either lower-than-expected opacity in the EBL or new physics \citep{lh23b, Galanti2023}.  }

{Knowledge of the redshift also provides robust constraints on the total neutrino luminosity for the GRB. Intensive study of candidate neutrino tracks in the IceCube experiment provided no evidence for any neutrino excess in the MeV, TeV, and PeV energy ranges \cite{icecube23}. Indeed, because of the proximity and high-energy emission (the processes for which can also create neutrinos), constraints on neutrino production for GRB 221009A are as strong as those from stacked analyses of larger numbers of bursts \citep{murase23}, and begin to provide meaningful constraints on at least some physical parameters within the fireball (e.g., bulk Lorentz factor and  dissipation radius; see \citealt{murase23,icecube23}).  }

\section{Conclusions}
We present a measurement of the redshift of GRB\,221009A with the VLT/X-shooter. Our observations allow us to measure $z=\zgrb$, demonstrating that GRB\,221009A is not only the brightest GRB ever seen (in terms of flux and fluence) but is also intrinsically one of the most energetic.

From the fluxes of a few emission lines detected on top of the afterglow spectrum, we are also able to constrain the SFR, metallicity, and dust content of the host galaxy. The values found are consistent with those of GRB host galaxies at low redshift.

In addition to the large tabulated amount of extinction in the direction of the GRB due to the MW ($A_V = 4.177$~mag), we detect extra absorption (by $\approx 0.5$~mag). From the shape of the afterglow SED, the fluxes of the host emission lines, and the detection of \ion{Na}{I} in absorption, we suggest that about $0.15$~mag of extinction is located in the GRB host, with the remaining lying in the MW.

Our analysis suggests that a burst this close and this luminous should only be witnessed at best once every few decades, and may be as rare as a once-per-millennium event, depending on its actual flux and fluence values.

\begin{acknowledgements}

We dedicate this paper to the memory of our beloved colleagues David Alexander ``Alex'' Kann, who was especially fond of this unique GRB.

We are grateful for a constructive report from an anonymous referee. This paper is based on observations collected at the European Southern Observatory under ESO programme 110.24CF (PI Tanvir). We thank the support from the ESO observing staff in Paranal, in particular Cédric Ledoux, Matias Jones, Michael Abdul-Masih, Zahed Wahhaj.

PDA acknowledge funding from the Italian Space Agency, contract ASI/INAF {n.} I/004/11/4 and from PRIN-MIUR 2017 (grant 20179ZF5KS).
DAK acknowledges the support by the State of Hessen within the Research Cluster ELEMENTS (Project ID 500/10.006).
DBM, AJL and NRT are supported by the European Research Council (ERC) under the European Union’s Horizon 2020 research and innovation programme (grant agreement No.~725246). The Cosmic Dawn Center is supported by the Danish National Research Foundation.
JH and LI were supported by a VILLUM FONDEN Investigator grant (project number 16599).
EPi acknowledges financial support through an INAF research grant.
MER acknowledges support from the research programme Athena with project number 184.034.002, which is financed by the Dutch Research Council (NWO).
AS and SDV acknowledge support from DIM-ACAV+.
KW acknowledges support through a UK Research and Innovation Future Leaders Fellowship awarded to Dr.~B.~Simmons (MR/T044136/1).
DX acknowledges the science research grants from the China Manned Space Project with NO. CMS-CSST-2021-A13 and CMS-CSST-2021-B11. 

\end{acknowledgements}

\bibliographystyle{aasjournal} 
\bibliography{refs} 

\end{document}